# Recent developments in Electromechanical Probing on the Nanoscale: Vector and Spectroscopic Imaging, Resolution, and Molecular Orientation Mapping

Sergei V. Kalinin,[*] S. Jesse,[*] A. Y. Borisevich,[*] H.N. Lee,[*] B.J. Rodriguez,[*]
J. Hanson,[†] A. Gruverman,[†] E. Karapetian,[#] and M. Kachanov[**]

[*] Oak Ridge National Laboratory,
Fax: 81-865-574-4143, e-mail: sergei2@ornl.gov
[†] North Carolina State University,
Fax: 81-919-515-3027, e-mail: Alexei_Gruverman@ncsu.edu
[#] Suffolk Universtiy,
Fax: 81-617-573-8591, e-mail: karap@comcast.net
[**] Tufts University
Fax: 81-617-627-3058, e-mail: Mark.Kachanov@tufts.edu

ABSTRACT: Strong coupling between electrical and mechanical phenomena and the presence of switchable polarization have enabled applications of ferroelectric materials for nonvolatile memories (FeRAM), data storage, and ferroelectric lithography. Understanding the local functionality of inorganic ferroelectrics including crystallographic orientation, piezoresponse, elasticity, and mechanisms for polarization switching, requires probing material structure and properties on the level of a single ferroelectric grain or domain. Here, I present recent studies on electromechanical, mechanical, and spectroscopic characterization of ferroelectric materials by Scanning Probe Microscopy. Three-dimensional electromechanical imaging, referred to as Vector Piezoresponse Force Microscopy, is presented. Nanoelectromechanics of PFM, including the structure of coupled electroelastic fields and tip-surface contact mechanics, is analyzed. This establishes a complete continuum mechanics description of the PFM and Atomic Force Acoustic Microscopy imaging mechanisms. Mechanism for local polarization switching is analyzed. The hysteresis loop shape is shown to be determined by the formation of the transient domain below the tip, the size of which increases with the tip bias. Spectroscopic imaging that allows relevant characteristics of switching process, such as imprint bias, pinning strength, remanent and saturation response, is introduced. Finally, resolution in PFM and vector PFM imaging of local crystallographic and molecular orientation and disorder is introduced.

## 1. INTRODUCTION

Recent progress in ferroelectric thin film technology and device manufacturing for applications such as non-volatile memories (FeRAMs), ferroelectric data storage, and ferroelectric lithography has necessitated the development of techniques for imaging and characterization of ferroelectric materials and devices on the nanoscale. In the last decade, Piezoresponse Force Microscopy (PFM) has emerged as a powerful tool for the ferroelectric domain structure imaging with sub-10 nm resolution. The spectroscopic modes of PFM, i.e. electromechanical hysteresis loop measurements, have also been demonstrated. Finally, PFM has been used to manipulate polarization locally, allowing domains with 20-50 nm size to be created, opening the pathway for high density data storage and ferroelectric lithography.

This tremendous progress is reflected in rapidly growing number of publications (> 120 papers/year), even though the vast majority of PFM studies are limited to qualitative observations of domain morphology and its evolution during phase transitions, switching, and fatigue. Here, we present several recent developments in PFM characterization of ferroelectric materials for quantitative studies of image formation mechanism, domain patterning, and switching phenomena. Theoretical developments include contact electromechanics of tip-surface junction, hysteresis loop shape and contrast transfer theory. Experimentally, we discuss vector PFM imaging, PFM resolution, and PFM image reconstruction. Finally, we discuss spectroscopic PFM imaging for spatially resolved mapping of switching characteristics of ferroelectrics, including imprint and coercive biases, work of switching, and remanent response. Several future directions for electromechanical characterization of nanoscale and disordered materials including polymers and biopolymers are discussed.

## 2. PRINCIPLES OF PFM

Piezoresponse force microscopy is based on the detection of the bias-induced piezoelectric surface deformation. The tip is brought into contact with the surface, and the piezoelectric response of the surface is detected as the first harmonic component, $A_{1\omega}$, of the tip deflection, $A = A_0 + A_{1\omega}\cos(\omega t + \varphi)$, induced by the application of the periodic bias, $V_{tip} = V_{dc} + V_{ac}\cos(\omega t)$, to the tip. Here, the deflection amplitude, $A_{1\omega}$, is assumed to be calibrated and given in the units of length. When applied to the pyroelectric or ferroelectric materials, the phase of the electromechanical response of the surface, $\varphi$, yields information on the polarization direction below the tip.

Application of the bias to the tip results in the surface displacement, **w**, with both normal and in-plane components, $\mathbf{w} = (w_1, w_2, w_3)$. The normal displacement of the tip apex in contact with the surface is generally equal to the surface displacement, since the effective spring constant of the tip-surface junction is typically 2-3 orders of magnitude higher than the cantilever spring constant. However, the lateral piezoresponse component in the direction normal to the cantilever axis (lateral transversal displacement), determined as torque of the cantilever, can be significantly smaller than that of the surface, e.g. due to the onset of sliding friction. Finally, typically ignored is the longitudinal surface displacement along the cantilever axis that couples to the vertical signal. The dynamic behavior of the cantilever and frequency dependence of these signal contributions are discussed in details elsewhere.[1,2]

## 3. EXPERIMENTAL

PFM, Atomic Force Acoustic Microscopy (AFAM) and Switching Spectroscopy Mapping PFM (SS-PFM) are implemented on a commercial Scanning Probe Microscopy system (Veeco MultiMode NS-IIIA) at ORNL equipped with additional function generators and lock-in amplifiers (DS 345 and SRS 830, Stanford Research Instruments, and Model 7280, Signal Recovery). A custom-built shielded sample holder was used to allow direct tip biasing in PFM and to avoid capacitive cross-talk in the SPM electronics. Measurements were performed using a variety of Pt and Au coated tips (e.g. NCSC-12 C, Micromasch, $l = 130$ μm, resonant frequency ~ 150 kHz, spring constant $k \sim 4.5$ N/m). Vertical PFM (VPFM) measurements were performed at frequencies 20 kHz – 2 MHz, which minimizes the longitudinal contribution to measured vertical signal.[1] In SS-PFM, optimal signal-to-noise ration was achieved at frequencies corresponding to the contact resonances of the cantilever. For lateral PFM (LPFM), the optimal conditions for contrast transfer were ~10 kHz; for higher frequencies, the onset of sliding friction minimizes in-plane oscillation transfer between the tip and the surface.[2,3] Custom LabView software was developed for simultaneous acquisition of VPFM, LPFM and AFAM phase and amplitude data, emulating additional SPM data acquisitions channels.

For AFAM and simultaneous AFAM-PFM measurements, the samples were glued to a commercial lead zirconium titanate (PZT) oscillator. To minimize cross-talk between PFM and AFAM signals, the top electrode was always grounded and a modulation bias at frequency $\omega_2$, was applied to the bottom electrode. The PFM modulation was at frequency $\omega_1$. Frequencies $\omega_1$ and $\omega_2$ are selected such that to avoid the overlap between higher and lower overtones. The output amplitudes, $A_n$, and phase shift, $\theta_n$, where $n = 1,2$ corresponds electrical to and mechanical excitations, respectively, are recorded by the SPM system electronics and ancillary computer.

To implement SS-PFM measurements, the microscope is configured similarly to force volume mode. The tip approaches surface vertically in contact mode until the deflection setpoint is achieved, and electromechanical hysteresis loop is acquired. Subsequently, the tip is moved to the next location so that $M \times M$ point mesh with spacing, $l$, between points is scanned. The hysteresis curves are collected in each point as a 3D data array and can be analyzed individually. Alternatively, parameters of switching process such as positive and negative coercive bias, imprint voltage, saturation response, and work of switching, can be plotted as 2D maps.

## 4. NANOMECHANICS OF TIP-SURFACE JUNCTION

Physical underpinnings of Scanning Probe Microscopy techniques can be conveniently understood using force-distance curves. Similar approach can be used for voltage modulation techniques such as PFM. However, here the system is described by two independent variables – tip-surface separation and tip bias, giving rise to force-distance-bias surface $F_c = F_c(h, V_{tip})$, where $h$ is indentation depth. Image formation mechanism in various SPM can be related to the derivatives of this surface, e.g. in the small signal approximation PFM signal is given by $(\partial h/\partial V_{tip})_{F=const}$ and AFAM signal is related to $(\partial h/\partial F)_{V=const}$. The rigorous solution of piezoelectric indentation, e.g. the shape of force-distance-bias surface, is available only for the case of transversally isotropic material.[4,5,6] Kalinin et al.[7] and Karapetian et al.[8] derived rigorous description of contact mechanics in terms of stiffness relations between applied force, $P$, and concentrated charge, $Q$, with indenter displacement, $w_0$, indenter potential, $\psi_0$, indenter geometry and materials properties. The solutions were obtained for flat, spherical, and conical indenter geometries, and have the following generalized structure:

$$P = \frac{2}{\pi}\theta\left(h^{n+1}C_1^* + (n+1)h^n\psi_0 C_3^*\right) \quad (1)$$

$$Q = \frac{2}{\pi}\theta\left(-h^{n+1}C_3^* + (n+1)h^n\psi_0 C_4^*\right) \quad (2)$$

where $h$ is total indenter displacement and $\theta$ is geometric factor ($\theta = a$ for flat indenter, $\theta = (2/3)R^{1/2}$ for spherical indenters and $\theta = (1/\pi)\tan\alpha$ for conical indenter) and $n = 0$ for flat, $n = 1/2$ for the spherical and $n = 1$ for the conical indenters, respectively.

These stiffness relations provide an extension of the corresponding results of Hertzian mechanics and continuum electrostatics to the transversely isotropic piezoelectric medium. All indentation stiffnesses are complex functions of electroelastic constants of material, $C_i^* = C_i^*(c_{ij}, e_{ij}, \varepsilon_{ij})$, where $c_{ij}$ are elastic stiffnesses, $e_{ij}$ are piezoelectric constants, and $\varepsilon_{ij}$ are dielectric constants. Detailed analysis of stiffness relations and effect of materials constants on values of coupling coefficients is given elsewhere.[7] It has been shown that for most materials $C_3^*/C_1^* \approx d_{33}$ and $C_4^* \approx \sqrt{\varepsilon_{11}\varepsilon_{33}}$, validating earlier approximations in PFM.

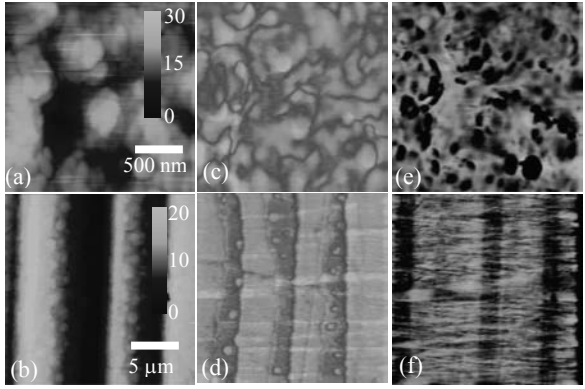

**Fig. 1.** Surface topography (a,b), PFM amplitude (c,d), and AFAM amplitude (e,f) for polycrystalline PZT thin film (a,c,e) and BaTiO$_3$ (100) surface (b,d,f). While PFM signal is relatively insensitive to contact area and hence topography, AFAM signal has significant contribution from topographic cross-talk.

Eqs. (1,2) yield a number of important conclusions on the information that can be obtained from SPM or nanoindentation experiment on the transversally isotropic piezoelectric material (e.g., $c^+$, $c^-$ domains in tetragonal perovskites). For all simple tip geometries, materials properties are described by three parameters, indentation elastic stiffness, $C_1^*$, indentation piezocoefficient, $C_3^*$, and indentation dielectric constant, $C_4^*$. Thus, the maximum information on electroelastic properties for a transversally isotropic material that can be obtained from an SPM experiment is given by these three quantities and mapping of $C_i^*$ distributions provides a comprehensive image of surface electroelastic properties. Experimentally, PFM signal is $d_{eff} = \partial h/\partial V_{tip} = C_3^*/C_1^*$. AFAM signal is related to the stiffness of tip surface junction, $k_1 = \partial P/\partial h = 2(n+1)\theta h^n C_1^*/\pi$. Due to the smallness of corresponding capacitance, indentation dielectric constant, $C_4^*$, can not be directly determined in the SPM experiment; however, it might be accessible on the larger length scales e.g. using nanoindentation approach.

From this analysis, PFM signal does not depend on tip-surface contact area and effective tip geometry, thus allowing for quantitative imaging. On the contrary, AFAM signal is strongly dependent on tip geometry, necessitating complex calibration procedures. Another implication is that PFM is only weakly dependent on surface topography, whereas AFAM will exhibit strong topographic cross-talk, as illustrated in Fig. 1 for PZT thin film (strong cross-talk) and BaTiO$_3$ (100) surface (nearly flat surface, the AFAM contrast is due to the Young modulus variation between $a$ and $c$ domains).

While the rigorous (and even approximate) description of piezoelectric indentation is not yet available for materials with lower symmetry, it can be conjectured that in analogy with indentation anisotropic elastic materials, Eqs. (1,2) will be valid for arbitrary materials. In addition, the in-plane component of surface displacement in this case will be non-zero. The zero order approximation for in-plane response is that components of displacement are given by the normal and shear elements of piezoelectric constant tensor, $(w_1, w_2, w_3) = (d_{34}, d_{35}, d_{33})$.[2]

## 5. VECTOR PFM

In general case, electromechanical response of the surface is a vector having three independent components. Standard beam deflection sensors allow quantitative measurements of two components, provided that the vertical and lateral sensitivities are properly calibrated using external (oscillator with zero frequency dispersion) or intrinsic (e.g. material with known responses and $a$-$c$ domain structures) calibration standard. In this case, PFM data can be represented as a 2D PFM vector image. An example of 2D PFM image is illustrated in Fig. 2, which shows vertical and lateral PFM images of PMN-PT single crystal. Vertical PFM image illustrates the presence of antiparallel $c$ domains, while lateral image shows the response at the domain walls due to tilting of the surface. To represent vector PFM data, the VPFM and LPFM images are normalized so that the intensity changes between -1 and 1, i.e. $vpr, lpr \in (-1,1)$. Using commercial software,[9] 2D vector data $(vpr, lpr)$ is converted to the amplitude/angle pair, $A_{2D} = \text{Abs}(vpr + I\, lpr)$, $\theta_{2D} = \text{Arg}(vpr + I\, lpr)$. This information can be represented using vector image, where the color corresponds to the orientation, while intensity corresponds to the magnitude. Alternatively, this data can be represented in the scalar form by plotting separately phase $\theta_{2D}$, and magnitude, $A_{2D}$, as illustrated in Figs. 2(d) and (e), respectively. Note that two types of domain walls can be observed on the amplitude image – "bright" walls parallel to the cantilever axis at which the electromechanical activity of the surface is enhanced, and "dark" walls perpendicular to the cantilever axis at which electromechanical activity is decreased. This asymmetry is due to the difference in signal transduction between longitudinal and lateral response components.

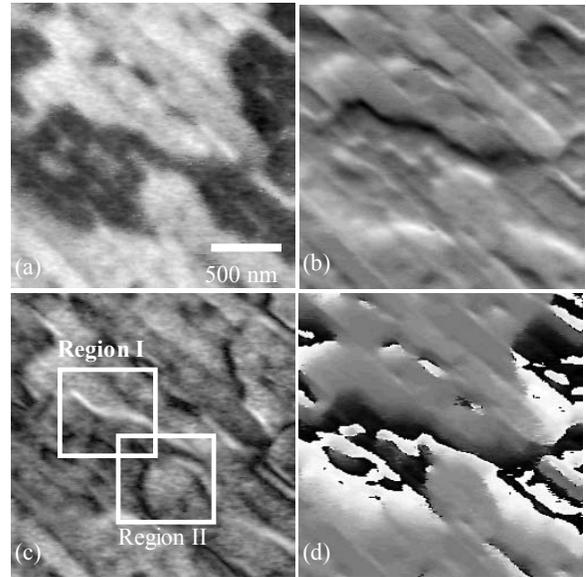

**Fig. 2.** (a) Vertical and (b) lateral PFM on PMN-PT surface illustrating the presence of antiparallel $c$ domains. The lateral contrast is due to surface tilt in the vicinity of domain walls. Corresponding (c) amplitude and (d) angle image. Note that depending on orientation domains walls are "bright" (in-plane response in lateral direction) and "dark" (in-plane response in longitudinal direction). This asymmetry is due to the difference in signal transduction between lateral and longitudinal components of surface displacement.

The third component of response vector can be obtained by physical rotation of sample by 90° and acquisition of PFM data from the same region. The details of data acquisition, interpretation and representation in this case are described in detail elsewhere.[2]

## 6. SWITCHING SPECTROSCOPY MAPPING IN PFM

Applications based on the switchable polarization necessitate the studies of fundamental mechanisms of domain switching and the influence of local defects such as grain boundaries and dislocations on switching behavior. This will ultimately allow fundamental mechanisms of imprint, fatigue, and retention in ferroelectric materials that determine their applicability for electronic applications, to be understood. On the macroscopic level, the vast amount of experimental and theoretical studies relating the switching characteristics to processing conditions, microstructure, etc. are available. Typically, information on the local switching processes, including imprint voltage, coercive bias, saturation response, and irreversible work of switching (energy loss), is obtained from hysteresis loops.

On the nanometer level, the information has been sparser. Recent studies by Gruverman et al.[10] has shown that domain nucleation in ferroelectric capacitors during repetitive switching cycles is always initiated at the same defect spots; similarly, the grain boundaries were shown to play important role in domain wall pinning.[11] Finally, domain freezing during fatigue has also been observed.[12] However, most studies to date have been limited to qualitative observations, i.e. no quantitative measures of local switching behavior have been obtained due to inherent limitations of PFM in imaging mode. Similarly to macroscopic case, PFM hysteresis loops will contain a wealth of information on local materials properties; however, the formation mechanism of electromechanical hysteresis loops in PFM are relatively unstudied and experimentally the information is collected in a few points along the sample surface, thus precluding correlation between materials microstructure and switching characteristics to be established. Here, we develop Switching Spectroscopy Mapping PFM (SS-PFM) to address switching characteristics of ferroelectric materials quantitatively on the nanometer level.

Shown in Fig. 3 is the idealized PFM hysteresis loop in the absence of electrostatic tip-surface interaction and electrostrictive coupling in the material. In this case, for high bias the material exists in the single domain state and response is bias independent. Of interest for applications are saturation response, $R_s^+$ and $R_s^-$, remanent response, $R_0^+$ and $R_0^-$, coercive biase, $V^+$ and $V^-$. Additionally, we define the critical bias corresponding to onset of switching, $V_c^+$ and $V_c^-$. From these quantities, we derive the imprint voltage, $I = (V^+ - V^-)$, and area of the loop (work of

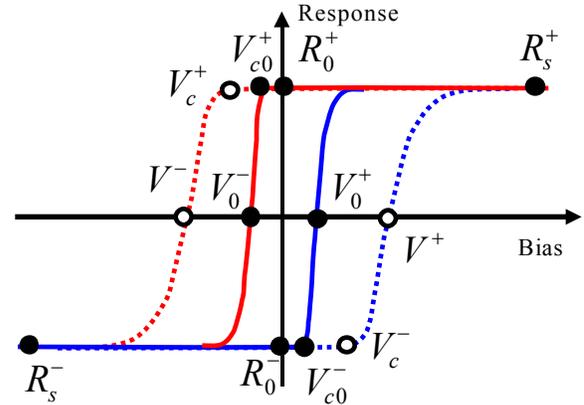

**Fig. 3.** Idealized electromechanical hysteresis loop in PFM in the thermodynamic (solid) and kinetic (dotted) regimes. Corresponding parameters are described in text.

switching), as

$$A_s = \int_{-\infty}^{+\infty} \left( R^+(V) - R^-(V) \right) dV, \quad (3)$$

where $R^+(V)$ is forward and $R^-(V)$ is reverse branch of the hysteresis loop. Experimentally, these quantities are determined either by statistical analysis of the data as functions of first and second moments of auxiliary function $\Delta^-(V) = R^+(V) - R^-(V)$, or through direct curve fitting by appropriate function. Data analysis and corrections for electrostatic interaction are described in detail elsewhere.[13]

SS-PFM imaging on a model polycrystalline PZT surface is illustrated in Fig. 4. Shown in Fig. 4(a) is the PFM image of polycrystalline PZT ceramics, illustrating the domain structure in the vicinity of the grain boundary (dotted line). The image is obtained form the analysis of the 2500 hysteresis loops obtained within 2 μm scan area, with the total acquisition time of 3 hours. The discontinuity in PFM image is associated with the change in grain orientation and hence possible domain orientations. Shown in Fig. 4(b) is the effective work of switching (area under hysteresis loop) from the same region. Note that the grain boundary is clearly associated with higher contrast, illustrating that energy loss during switching process is increased by a factor of ~2 compared to the grain bulk. Figure 4(c) and (d) illustrate the imprint map and hysteresis loop width, correspondingly. Note that there is no systematic correlation between PFM and these maps, which thus provide complementary information to conventional PFM.

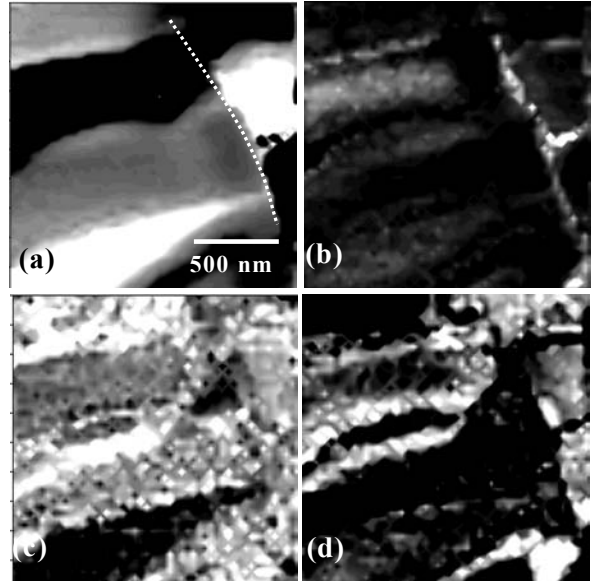

**Fig. 4.** (a) Piezoresponse image, (b) work of switching, (c) imprint voltage and (d) and hysteresis loop width maps for polycrystalline PZT ceramics. Scale is (c) 0.5 – 3.6 V and (d) 2.2-3.8 V.

To relate the hysteresis loop parameters to materials properties, we introduce the following simple model. In PFM, the electric field is concentrated directly below the tip, resulting in preferential domain nucleation at the tip-surface junction. In thin films, this nascent domain becomes stable when the domain length, $l$, becomes equal to the film thickness, $h$, in which case the depolarization energy decreases due to the effective polarization charge compensation by the back electrode. In crystals, the stability is achieved when restoring force due to the depolarization field and domain wall energy is insufficient to overcome pinning by lattice (intrinsic) and defects (extrinsic). The shape of the hysteresis loop in PFM was analyzed by Kalinin et al.[14] using 1D model by Ganpule.[15] It was conjectured that $PR = d_{eff}\{V(0) - 2V(l)\}$, where $V(0)$ is the potential on the surface and $V(l)$ is the potential at the domain boundary below the tip, and $d_{eff}$ is effective electromechanical response of material.[7] In the strong indentation regime, when the potential distribution inside the material can be approximated using point-charge model potential decays as $V(l) = \beta V(0)a/l$, where $\beta$ is proportionality coefficient of the order of unity, $l$ is the distance from the center of the contact area and $a$ is contact radius. Thus, the PFM loop shape is controlled by the shape of the nascent domain below the tip as

$$PR = d_{eff}\{1 - 2\beta a/l(V_{dc})\}. \qquad (4)$$

In the point charge approximation applicable for the late stages of domain growth, the equilibrium domain length is $l = d/5b$, where $b = \sigma_{wall}\pi^2/2$ is determined by direction-independent domain wall energy, $\sigma_{wall}$, and $d = d_0 V_{dc} = 2P_s C_c V_{dc}/(\varepsilon_0 + \sqrt{\varepsilon_{11}\varepsilon_{33}})$, where $P_s$ is polarization, and $\sqrt{\varepsilon_{11}\varepsilon_{33}}$ is effective dielectric constant for transversally-isotropic material.[16,17] In the strong indentation limit, the capacitance of tip-surface junction is $C_c \approx 4a\sqrt{\varepsilon_{11}\varepsilon_{33}}$, thus yielding $d_0 \approx 8aP_s$ (this also holds for field generated by spherical part of the tip). Hence, the shape of the PFM hysteresis loop for large bias voltages is expected to follow functional form

$$PR = d_{eff}\{1 - 5\beta\sigma_{wall}\pi^2/(8P_s V_{dc})\}. \qquad (5)$$

In this thermodynamic limit, the shape of the hysteresis loop is independent on dielectric properties of material and the tip-surface contact radius and is determined solely by the domain wall energy and spontaneous polarization. Note that in the complete absence of pinning, the thermodynamic model suggests that the hysteresis will proceed along the positive or negative branch in Fig. 3, i.e. process is reversible (domain grows and shrinks in response to tip bias). However, with the small pinning, the domain wall movement below the tip can be expected to be much faster than on the outer domain boundary, resulting in the characteristic hysteretic behavior.

The saturation and remanent responses in this approximation are equal, and the onset of switching corresponds to the domain nucleation below the tip. Critical bias $V_c^+ = V_c^- = 0$ in the point charge limit when the field below tip is infinite, or small finite value for finite electric field corresponding to experimental conditions. In the latter case, a small hysteresis corresponding to critical bias for domain nucleation will be observed, as expected for the first-order phase transition.[18] The imprint voltage is determined by the contact conditions at the tip-surface junction and build-in field in the material and can not be determined in this model.

From Eq. (5), the coercive bias in the thermodynamic limit can be derived as

$$V_0^+ = V_0^- = 5\beta\sigma_{wall}\pi^2/8P_s \qquad (6)$$

For BaTiO$_3$ ($\sigma$ = 7 mJ/m$^2$, $P_s$ = 0.26 C/m$^2$)[19] the estimated coercive bias is $V^+$ = 0.166 V (for $\beta$ = 1). Note that while Eqs. (4-6) are derived in the 1D approximation, in general case the effective piezoresponse is zero for some fixed domain size, $l_{crit}$, for which the response from the nascent domain below the tip and the surrounding material are compensating each other, giving rise to fundamentally similar behavior.

It has been recently shown that the domain growth in ferroelectric materials is kinetically limited process and domain wall velocity is determined by the local field and presence of pinning centers (lattice and defects).[20,21] In the presence of pinning, hysteresis loop will broaden compared to the thermodynamic limit, as shown in Fig. 3. Note that the saturation response corresponding to complete switching is not affected by the pinning.

In this kinetically limited case, the hysteresis loop shape can be described as following. Experimentally, domain switching in ferroelectric materials has been shown to follow the phenomenological dependence $r(V,t) = Vg(t)$, where $g(t)$ is the universal scaling function.[21] For small times, $g(t) \sim \ln t$. Thus, the kinetics of domain switching for small domain sizes is $r(V,t) = k_s V \ln t$ and domain wall velocity is $\dot{r} = k_s V/t$, where $k_s$ is the kinetic constant related to the pinning strength in the material. Note that while the kinetic data was obtained for the lateral domain size, the time dependences for vertical and lateral sizes can be expected to be commensurate.

In the hysteresis measurements, the tip bias is ramped linearly with time, $V = bt$. Hence, $\dot{r} = k_s b$ and $r = k_s bt = k_s V$. Therefore, the domain size is independent of the ramp velocity and is determined by the pinning strength in material only. Notably, this is also the case for general power-law ramps, $V \sim t^n$, $n > 0$. Thus, we recover the functional form of Eq. (5), which thus provides a robust measure of hysteresis loop shape. In the kinetically limited case, the coercive bias is now $V^+ = V^- = 2\beta a/k_s$, and is controlled by the pinning in the material.

Finally, to extend this analysis to the experimental data, we analyze additional factors influencing loop shape. The hysteresis loop broadening can occur due to imperfect tip-surface contact, i.e. dielectric gap, resulting in the attenuation of effective surface potential, $V(0) = \alpha V_{tip}$.[22] This gap effect will also result in the decrease of electromechanical response, $PR = d_{eff}/\alpha$. Noteworthy is that for frequency-independent dielectric constant of the gap, $\varepsilon_{dc} = \varepsilon_\omega$, the product of coercive bias and saturated electromechanical response is independent on gap effect, $R^+V^+ = R_0^+V_0^+ = const(\alpha)$. Thus, the *area* of the hysteresis loop (rather than loop width) provides a robust measure of the pinning strength in the material. Also note that the electrostatic and electrostrictive contributions to the PFM signal are conservative (for linear elastic material) and thus do not contribute to the area below the loop.

To summarize, SS-PFM allows real space imaging of switching properties of ferroelectric materials such as imprint bias, saturation response, etc. In particular, the area below the hysteresis loop is a measure of the pinning strength and hence losses in materials, and provides important information for device applications.

## 7. RESOLUTION AND IMAGE RECONSTRUCTION

Applications of Piezoresponse Force Microscopy for ferroelectric data storage and ferroelectric lithography necessitate the resolution and minimum writable domain size in PFM and other ferroelectric domain imaging techniques to be established. To date, several claims of sub-10 nm (in PFM)[23,24] and even sub-nm resolution (in SNDM)[25] imaging have been made based solely on the minimum observed feature size. Reports on domain sizes achievable by high density ferroelectric recording (as small as 40 nm,[20] 20 nm[26] and even 12 nm[27]) are similarly based. The further progress in the field necessitates the consistent definition for *resolution* and *minimal observable feature size* in PFM to be established, to compare the results obtained by different groups and different techniques, determine the veracity of the PFM data storage and ferroelectric lithography, and ultimately deconvolute materials properties and probe effect using transfer function approach.

The analysis of image formation mechanism in SPM is greatly simplified if the imaging is linear. In this case, the measured image, $I(\mathbf{x})$, is a convolution between ideal image, $I_0(\mathbf{x}-\mathbf{y})$, representing material (sample) parameters and the instrument (probe) function, $F(\mathbf{y})$:

$$I(\mathbf{x}) = \int I_0(\mathbf{x}-\mathbf{y})F(\mathbf{y})d\mathbf{y} + N(\mathbf{x}) \qquad (7)$$

where $N(\mathbf{x})$ is the noise function. The linear theory has been shown to be applicable for PFM imaging for (a) weak electromechanical coupling (typical error for material such as $BaTiO_3$ or $LiNbO_3$ is 10-20%) and (b) electromechanical properties uniform in *z*-direction on the length scale of electric field penetration (e.g. 180° domain walls).[28] The Fourier transform of Eq. (5) is

$$I(\mathbf{q}) = I_0(\mathbf{q})F(\mathbf{q}) + N(\mathbf{q}) \qquad (8)$$

where $I(\mathbf{q}) = \int I(\mathbf{x})e^{i\mathbf{qx}}d\mathbf{x}$, $I_0(\mathbf{q})$, $F(\mathbf{q})$ and $N(\mathbf{q})$ are the Fourier transforms of the measured image, ideal image, microscope function and noise, respectively. The instrument function can then be determined directly provided that the ideal image, $I_0(\mathbf{q})$, is known. Thus, once the instrument function is determined for a known calibration standard, it can be used to evaluate the ideal image, $I_0(\mathbf{x})$, from the measured image, $I(\mathbf{x})$ for an arbitrary sample. The width of $F(\mathbf{y})$ provides a quantitative measure of resolution.

For PFM, resolution function $F(\mathbf{y})$ depends on tip geometry, contact conditions, etc, and its calculation from geometric parameters of the tip is subject to multiple uncertainties. However, from the linearity it can be also determined experimentally from a well defined calibration standard, e.g. artificially engineered domain pattern.

Illustrated in Fig. 5(a) is a voltage pattern used to create domains on the PZT surface, while corresponding Fourier transform (FT) is shown in Fig. 5(b). Note that "domain walls" in Fig. 5(a) are extremely sharp; hence all (*hk*) peaks can be seen on the FT. For this symmetric lattice, the extinction rule is $h-k = 2n+1$, $n = 0,1, ...$. In comparison, shown in Fig. 5(c,d) is resultant domain pattern and corresponding FT. Note that only several low order reflections can be observed on FT; also, the peaks which were zero intensity in written image are now seen (arrow). This reconstruction is due to the fact that the area of the "white" and "black" domains is different on the written image, due to significant imprint (~1-2 V) of the film.

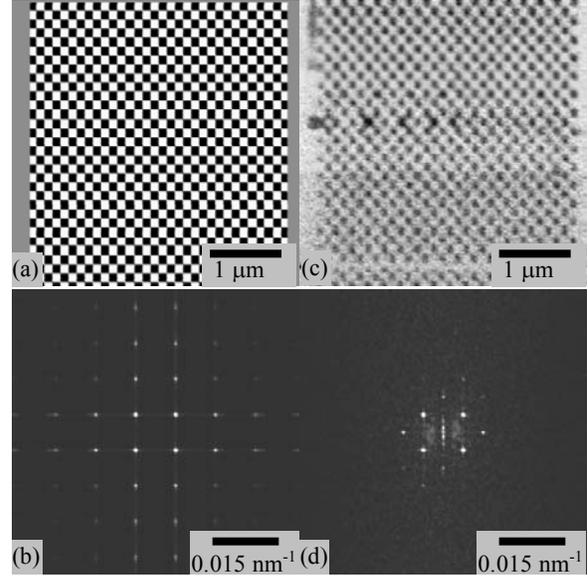

**Fig. 5.** (a) Ideal image (writing signal) and (b) corresponding FFT image illustrating that all frequency components are present. PFM images acquired with 1 ms lock-in time constants and (d) corresponding FFT image.

The position of the peak with highest $q$ (shown by the circle) which is still above noise background, $I(hk) > N(q)$, determines the minimal feature size in the PFM image. To quantify this behavior further, shown in Fig. 6(a) is the wavevector dependence of peak intensities. Experimentally, $I(hk) = I_0 \exp(-q/G)$ with $G \approx 0.011$ nm$^{-1}$ and

-6-

can be approximated by Gaussian $F(q) = A\exp(-q^2/2w^2)$, where $A = (1.45 \pm 0.08)\, 10^{-3}$ and $w = (12.2 \pm 0.8)\, 10^{-3}$ nm$^{-1}$. *Resolution* can now be (somewhat arbitrary) defined as $1/q$ for which $F(q) = 0.1 F(0)$. In this particular case, resolution is ~40 nm and close to minimal feature size. Despite this closeness of numerical values, note that the fundamental difference between the resolution and minimal feature size is that the former is a universal characteristic of the microscope, whereas the latter is determined by the noise level in the system.

The complementary real space definition of minimal feature size is illustrated in Fig. 7, where the domain pattern is created using the variable size mesh. The profile along the dotted line is illustrated in Fig. 7(d). Note that the domain wall width is ~15 nm, providing an (not unambiguous) intrinsic measure of the resolution of the technique. For large domains, signal saturates both on top and bottom of the domain, while for smaller domains the domain walls overlap, resulting in characteristic triangular shape. Thus, *minimal detectable feature size* can be significantly smaller than domain wall width (resolution) and is determined both by the resolution and the noise level of the microscope. Also, unlike Fourier criterion in Fig. 6, the determination of minimal visible feature size in real space can be subject to multiple uncertainties.

For linear imaging, the experimentally determined resolution function can be used to reconstruct the "ideal image", as demonstrated in Fig. 7(c). The written pattern and corresponding domain pattern are shown in Fig. 7(a,b). For deconvolution, the recorded image FT was divided by the resolution function. To avoid the spurious amplification of the large-$q$ features, the noise offset was selected as $0.1 F(0)$. The resulting deconvoluted image is shown in Fig. 7(c). Note the difference in the image contrast. This behavior is further illustrated in Fig. 7(d,e), showing line profiles across the images in Fig. 7(b,c). Note the decrease in domain wall width, indicative of ideal image. Also note that the minimal domain size detected by PFM in this case is limited by the resolution of the technique, suggesting that the resolution is a limiting factor precluding experimental observation of smaller domains that can be created by PFM.

In general case of non-uniform field distribution PFM image can not be represented as a convolution Eq. (7), similarly to topographic AFM imaging.[29] In such cases, however, it may still be possible to analyze spatial frequencies revealed in the image through a Fourier transform (FT) provided that the features are real, as suggested for AFM by Engel[30] and Gutierrez.[31]

## 8. ORIENTATION AND DISORDER BY PFM

In addition to piezoelectric and ferroelectric domain imaging, quantitative local electromechanical measurements open at least two novel venues for characterization of materials nanostructures. Piezoelectricity is described by a rank 3 tensor, and is thus strongly orientation dependent. Thus, quantitative electromechanical measurements can provide information on local crystallographic orientation, i.e. relationship between the coordinate system linked to crystal and laboratory. The coordinate transformation between the two requires three rotations described by the Euler angles $\phi$, $\theta$, and $\psi$. For crystalline materials, the relationship between $q = \sqrt{h^2 + k^2}/a$, $a$ is periodicity of the lattice. The minimum detectable feature size is determined by condition $I(hk) = N(q)$ and from Fig. 6(a) can be estimated as ~33 nm. To calculate the transfer function in Eq. (8), shown in Fig. 6(b) is the ratio of the FT peak intensities for experimental and ideal images. The resulting dependence the piezoelectric constant tensor in the laboratory coordinate system, $d_{ij}$, and the tensor in the crystal coordinate system, $d_{ij}^0$, is:

$$d_{ij} = A_{ik} d_{kl}^0 N_{lj} \quad , \quad (9)$$

where the matrices $N_{ij}$ and $A_{ij}$ are functions of the Euler angles.[32] As an example, we consider tetragonal $BaTiO_3$. In the coordinate system of the crystal, the $d_{ij}^0$ tensor is

$$d_{ij}^0 = \begin{pmatrix} 0 & 0 & 0 & 0 & d_{15}^0 & 0 \\ 0 & 0 & 0 & d_{15}^0 & 0 & 0 \\ d_{31}^0 & d_{31}^0 & d_{33}^0 & 0 & 0 & 0 \end{pmatrix} \quad (10)$$

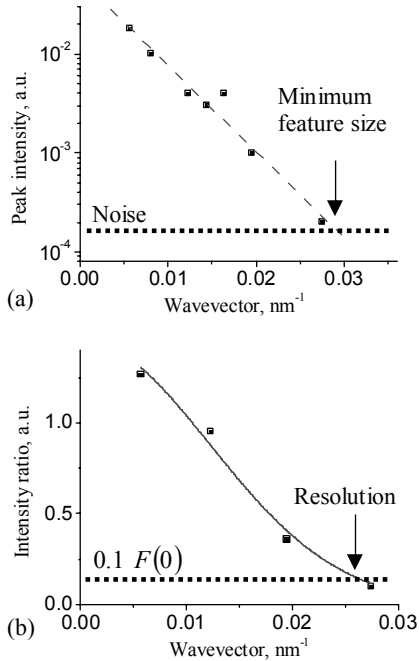

**Fig. 6.** (a) PFM image of the grid patter and (b) corresponding FFT image. (c) Wavevector dependence of the FFT peak intensity illustrating the minimal feature size. (d) Calculated transfer function, illustrating resolution.

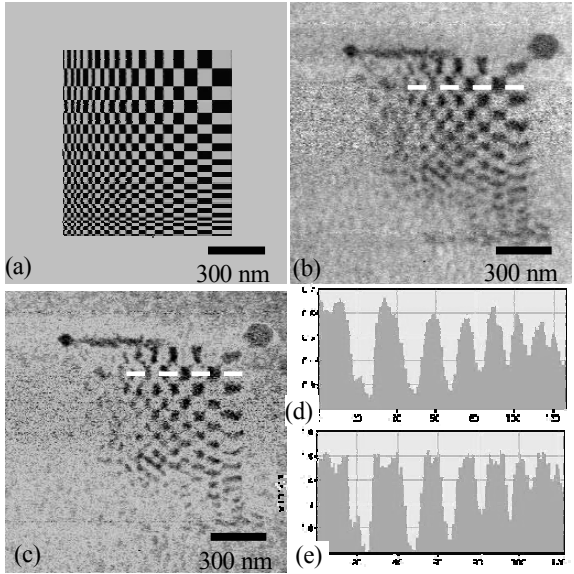

**Fig. 7.** (a) Writing pattern (b) original PFM image and (c) reconstructed PFM image using the transfer function. (d) original and (e) reconstructed profiles along the dotted lines in (b,c). Notice the difference in domain wall width.

For a general orientation of the crystal, the response components relevant to PFM are:

$$d_{33} = (d_{15}^0 + d_{31}^0)\sin^2\theta\cos\theta + d_{33}^0\cos^3\theta \quad (11)$$

$$d_{34} = -(d_{31}^0 - d_{33}^0 + (d_{15}^0 + d_{31}^0 - d_{33}^0)\cos 2\theta)\cos\psi\sin\theta \quad (12)$$

$$d_{35} = -(d_{31}^0 - d_{33}^0 + (d_{15}^0 + d_{31}^0 - d_{33}^0)\cos 2\theta)\sin\psi\sin\theta \quad (13)$$

This strong orientation dependence of electromechanical response provides an approach for mapping local crystallographic orientation, i.e. if the elements of the piezoelectric constant tensor can be experimentally measured, local crystallographic orientation, $(\phi_i, \theta_i, \psi_i)$, can be completely or partially derived.

Piezoelectric coupling can emerge even in the *partially* ordered polar materials, including poled ferroelectric ceramics, ferroelectric polymers, and many biological systems, such as connective and calcified tissues and wood. For such materials, piezoelectric coupling is directly related to degree of ordering. For example, for a texture of disordered tetragonal crystal with axial disorder, the effective piezoelectric coefficients for texture, $d_{ij}^0$, is related the piezoelectric coefficients for the original material, $d_{ij}^1$, as (for normal response coefficient):[33]

$$d_{33}^0 = \frac{1}{4}(1+\cos\theta_c)[(1+\cos^2\theta_c)d_{33}^1 + \sin^2\theta_c(d_{15}^1 + d_{31}^1)] \quad (14)$$

where $\theta_c$ is a parameter (angular distribution of crystallites) that describes the texture. For large disorder, the response decreases rapidly, becoming zero for isotropic system, $\theta_c = \pi$. Thus, measurement of electromechanical coupling provides a degree of ordering in material.

To summarize, in the last decade Piezoresponse Force Microscopy has evolved from nascent and controversial technique for ferroelectric domain imaging to powerful quantitative tool for imaging static and dynamic properties of ferroelectric and piezoelectric materials with nanometer resolution. The next decade will undoubtedly see new instrumental and theoretical developments in the rapidly developing field of nanoscale electromechanics and ferroelectric phenomena, to which SPM and particularly PFM provide the first and the only key.

Research supported by Oak Ridge National Laboratory, managed by UT-Battelle, LLC, for the U.S. Department of Energy under Contract DE-AC05-00OR22725.